\documentclass[12pt]{iopart}
\usepackage{graphicx}

\begin{document}
\title[Microscopic theory of large-amplitude collective motion]
{Open problems in microscopic theory of large-amplitude collective motion}

\author{K Matsuyanagi$^{1,2}$, M Matsuo$^3$, 
T Nakatsukasa$^1$,  \\ N Hinohara$^1$ and  K Sato$^{1,4}$}
\address{$^1$Theoretical  Nuclear Physics Laboratory, 
RIKEN Nishina Center, Wako 351-0198, Japan}
\address{$^2$Yukawa Institute for Theoretical Physics, Kyoto University, 
Kyoto 606-8502, Japan}
\address{$^3$Department of Physics, Faculty of Science, 
Niigata University, Niigata 950-2181, Japan}
\address{$^4$Department of Physics, Graduate School of Science, 
Kyoto University, Kyoto 606-8502, Japan} 

\begin{abstract}
Construction of the microscopic theory of large-amplitude collective motion,
capable of describing a wide variety of quantum collective phenomena 
in nuclei, is a long-standing and fundamental subject in the study of 
nuclear many-body systems.  
Present status of the challenge toward this goal is discussed 
taking the shape coexistence/mixing phenomena as typical manifestations of 
the large-amplitude collective motion at zero temperature.
Some open problems in rapidly rotating cold nuclei are also briefly discussed 
in this connection.
\end{abstract}
\pacs{21.60.Ev, 21.60.Jz}
\submitto{\JPG}
\newcommand{\bra}[1]{\left\langle #1 \right|}
\newcommand{\ket}[1]{\left| #1 \right\rangle}
\newcommand{\del}{\partial}
\newcommand{\Hhat}{\hat{H}}
\newcommand{\Nhat}{\hat{N}}
\newcommand{\Qhat}{\hat{Q}}
\newcommand{\Phat}{\hat{P}}
\newcommand{\Ghat}{\hat{G}}
\newcommand{\That}{\hat{\Theta}}
\newcommand{\Nt}{\tilde{N}}
\newcommand{\Hc}{{\cal H}}

\section{Introduction}

Low-frequency collective modes of excitation in cold nuclei near the yrast line  
exhibit a number of unique features of the nucleus as 
a finite quantum many-body system. 
To understand the nature of these collective excitations, we need to develop 
a microscopic theory of large-amplitude collective motion (LACM),  
which has sound theoretical basis and, 
at the same time, is practical enough for applications to  
a wide variety of nuclear collective phenomena. 
This is a very broad and long-term pursuit in nuclear structure physics. 
Through the attempts up to now toward constructing the microscopic theories of LACM,   
promising new concepts and methods have been proposed and developed 
(see \cite{dan00} for a review), but 
it may be fair to say that the challenge is still in its infancy.   
In this article, we discuss the present status of the challenge 
taking mainly the shape coexistence/mixing phenomena 
as typical manifestations of LACM at zero temperature.
Open problems in the theoretical formulation 
of microscopic LACM theory and collective phenomena awaiting its application 
are listed including those in rapidly rotating cold nuclei. 
This article is not intended to be a comprehensive review 
of this broad field of nuclear structure physics, and 
we apologize that the selection of topics and references is leaning to  
our personal interest.  
Certainly, microscopic description of spontaneous fission 
from the viewpoint of non-linear/non-equilibrium physics 
is one of the major goals, but this subject will be discussed by other contributors 
to this special issue on {\it open problems in nuclear structure}. 

\section{Shape coexistence/mixings as large-amplitude collective phenomena}

\subsection{Shape coexistence phenomena}
In recent years, new experimental data exhibiting  coexistence 
of different shapes (like spherical, prolate, oblate and triaxial shapes) 
in the same energy region (of the same nucleus)  have been obtained  
in low-energy spectra of nuclei in various regions of nuclear chart   
(see \cite{abe90,woo92} for reviews and \cite{and00,fis00,cle07}
for examples of recent data).
These data indicate that the shape coexistence is  
a universal phenomenon representing essential features of the nucleus. 

\subsection{Necessity of going beyond the small-amplitude approximation}
Let us define `shape of the nucleus'  as a semi-classical, macroscopic concept 
introduced by the self-consistent mean-field approximation, 
such as the Hartree-Fock-Bogoliubov (HFB) approximation, to 
the quantum-mechanical many-nucleon system. 
Needless to say, any HFB equilibrium shape inevitably accompanies  
quantum zero-point oscillations. 
If only one HFB equilibrium state exists, we can describe various kinds of 
vibrational mode about this point using the standard many-body methods 
like the random-phase approximation (RPA)  and 
the boson-expansion methods \cite{rin80, bla86}. 
In the situations where two different HFB equilibrium shapes coexist 
in the same energy region, however, the large-amplitude shape vibrations 
tunneling through the potential barrier between the 
two HFB local minima may take place. 
To describe such LACM, 
we need to go beyond the perturbative approaches based on an expansion 
about one of the local minima.   　

\subsection{Quantum field theory point of view}
Different from the well-known tunneling of a single-particle through 
the barrier created by an external field, the shape mixing of interest 
between different HFB local minima is a macroscopic tunneling phenomenon 
where the potential barrier itself is generated as a 
consequence of the dynamics of the self-bound quantum system.
A HFB local minimum corresponds to a vacuum for quasiparticles in 
the quantum field-theoretical formulation. 
In contrast to infinite systems, different vacua in a finite quantum system 
are not exactly orthogonal to each other.  
Thus,  the shape coexistence phenomena provide us precious opportunities 
to make a detailed study of the many-body dynamics of LACM  
connecting different vacua in terms of quantum spectra 
and electromagnetic transition properties associated with them.  

\subsection{Unique feature of the oblate-prolate coexistence}
When the self-consistent mean field breaks the reflection symmetry 
(like the pear shape), parity doublets appear as the symmetric and 
anti-symmetric superpositions of two degenerate states associated with 
the two local minima.  In contrast, 
there is no exact symmetry like parity
when two local minima having the oblate and prolate shapes coexist. 
In this case, we do not know the relevant collective degree(s) of freedom 
through which the two shapes mix. 
It is obviously needed to develop a microscopic theory capable of describing 
the shape mixing dynamics of this kind. 

\subsection{Mysterious $0^+$ states}
There are only a few nuclei in which the first excited  $0^+$ state appears 
below the first excited $2^+$ state. A well known example is the 
$0^+$ state at $0.69$ MeV in $^{72}$Ge which appears below the 
$2^+$ at $0.83$ MeV. The nature of this state is poorly understood.  
Experimental systematics including neighboring nuclei indicates that the 
excitation energy of the $0^+$ state takes the minimum
around $N=40$ where the neutron pairs starts to occupy the $g_{9/2}$ shell.  
Microscopic calculations \cite{tak86,wee81} using the boson expansion method  
indicate that the mode-mode coupling between the quadrupole anharmonic vibration  
and the neutron pairing vibration plays an indispensable role in bringing about the 
peculiar behavior of the excited $0^+$ states.  
On the other hand, these states are often interpreted in terms of the 
phenomenological shape coexistence picture \cite{kot90}. 
Relations between the two interpretations are not well understood.  
Closely examining the properties of the excited $0^+$ states 
in a wide region of nuclear chart, 
one finds that they exhibit features that cannot be understood 
in terms of the traditional concepts alone 
(see \cite{gar09} for an example of recent data). 

\section{Characteristics of low-frequency collective excitations in nuclei}

As is well known, shell structure and pairing correlations play  
essential roles for the emergence of low-frequency collective modes of excitation  
in medium-heavy and heavy nuclei. 
They exhibit unique features of the nucleus 
as a finite quantum many-body system and 
their amplitudes of vibration tend to become large.   
In this section, we discuss their characteristics in a wider perspective 
including the shape mixing phenomena  
and argue for the need of a microscopic theory capable of describing them. 

\subsection{Deformed shell structure}
Nuclei exhibit quite rich shell structures comprising of 
a variety of single-particle motions in the mean field localized in space.   
Let us define the shell structure as a regular oscillating pattern in the 
single-particle level density coarse-grained in energy. 
The nucleus gains an extra binding energy, called shell energy, 
when the level density at the Fermi surface is low. 
The shell structure changes as a function of deformation.
If the level density at the Fermi surface is high at the spherical shape, 
the nucleus prefers a deformed shape with lower level density. 
When different deformed shell structures give almost the same energy gain, 
we may obtain approximately degenerate HFB equilibrium shapes. 

\subsection{Pairing correlations and quasiparticles}
It should be emphasized that both the mean field and the single-particle modes 
are collective phenomena. Needless to say, 
the nuclear mean field is self-consistently generated 
as a result of cooperative motion of strongly interacting nucleons. 
We learned from the BCS theory of superconductivity 
that the single-particle picture emerges 
as a consequence of collective phenomena: 
the Bogoliubov quasiparticles are nothing but the elementary modes of excitation 
in the presence of the Cooper pair condensate, 
and they have a gap in their excitation spectra (become `massive'). 
As is well known, this idea has been greatly extended to understand 
dynamical mechanisms of generating the masses of `elementary' particles 
\cite{nam60}. 
The HFB theory is a generalized mean field theory 
taking into account both the pair condensation and the 
HF mean field in a unified manner \cite{rin80,bla86,ben03}.

\subsection{Spontaneous breaking of symmetry}
The self-consistent mean field of a finite quantum system inevitably 
breaks some symmetries.  Even the spherical mean field breaks the 
translational symmetry. When the mean field breaks another symmetry
of a higher order, the concept of single-particle motion is generalized 
accordingly. For instance, the Bogoliubov quasiparticle is   
introduced by breaking  the number conservation. 
Struggles for finding a better concept of single-particle motion 
comprise the heart of nuclear structure study. 
When the mean field breaks some continuous symmetries,  
collective modes (Nambu-Goldstone modes) restoring the broken 
symmetries emerge. 
Nuclear rotations are typical examples: 
they restore the rotational symmetries broken by the mean field \cite{boh75}. 
In this way, the (generalized) single-particle picture and 
the symmetry-restoring collective motions are inextricably linked like
`two sides of the same coin.'
This fact has been beautifully demonstrated by the success of the 
`rotating (cranked) shell model' \cite{fra01} which describe the 
interplay of the rotational motions and 
`the single-particle motions in the rotating mean field' in a simple manner. 
One of the fascinations of nuclear structure physics is that 
we can study the microscopic dynamics of symmetry breaking and 
restoration by means of a detailed study of quantum spectra. 
Finite quantum systems localized in space, such as the nucleus,  
provide us with such unique and invaluable opportunities. 

\subsection{Origin of oblate-prolate asymmetry}
It should be noted that breaking of the spherical symmetry does not necessarily 
lead to the regular rotational band structure.
For instance, even when the HFB mean field has an equilibrium point 
at a prolate shape which is deep with respect to 
the axial deformation parameter $\beta$, it should be deep also 
with respect to the axial asymmetric deformation parameter $\gamma$. 
In HFB calculations restricted to the axially symmetric shapes, 
we often obtain two solutions having the prolate and oblate shapes, 
but the oblate solution might be unstable with respect to $\gamma$. 
Even if both minima are stable, a strong mixing of the two shapes might 
occur through quantum mechanical large-amplitude collective vibrations 
in the $\gamma$ degree of freedom. 
To suppress such a mixing, we need a sufficient amount of the potential barrier
and the energy difference between the oblate and prolate solutions.   
Otherwise, identities of the rotational bands built on the oblate and prolate 
shapes will be easily lost due to the large-amplitude $\gamma$ vibrations. 
These problems point to the critical need for 
the microscopic theory of LACM capable of describing 
such $\gamma$-soft situations, the oblate-prolate shape coexistence
(where two rotational bands built on them can be identified)
and various intermediate situations in a unified manner.   
In spite of the obvious importance for understanding low-energy collective spectra, 
the microscopic origin of the oblate-prolate asymmetry in nuclear structure
(the reason why the prolate shapes are energetically more favored in many cases 
than the oblate shapes) still remains as 
one of the long-standing and fundamental problems.  
On this issue, quite recently,  Hamamoto and Mottelson suggested that the surface 
diffuseness plays a key role in bringing about the asymmetry \cite{ham09}.
A useful approach to investigate the dynamical origin of appearance of 
the deformed shell structure is the semi-classical theory of shell structure 
\cite{bra97,fri90,ari98}.
It may be interesting to apply this approach to the problem in question 
for a deeper understanding of the oblate-prolate asymmetry. 

\subsection{Collective motion as moving mean field} 

Nuclear rotational and vibrational motions can be described as 
moving self-consistent fields. 
This is one of the basic ideas of the unified model of 
Bohr and Mottelson \cite{boh76,mot76}. 
The time-dependent HFB (TDHFB) mean field is a generalized coherent state 
and its time development can be described as a trajectory in 
the large-dimensional TDHFB phase space.
Such a formulation of the TDHFB theory as a Hamilton dynamical system 
provides a microscopic foundation 
for using a classical picture of rotating and vibrating mean fields 
\cite{neg82,abe83,kur01}.
Thus, nuclear collective motions are beautiful examples of 
emergence of classical properties in genuine quantum many-body systems. 
For small-amplitude vibrations around a HFB equilibrium point, 
one can make the linear approximation to the TDHFB equations 
and obtain the quasiparticle RPA (QRPA). 
One of the merits of the QRPA is that 
it determines the microscopic structures of the normal modes (collective coordinates)
without postulating them from the outset. 
The small-amplitude approximation is valid for giant resonances 
(high-frequency collective vibrations), and the (Q)RPA   
is used as the standard method for their microscopic descriptions. 
  
\subsection{Need for a microscopic theory of LACM}

In contrast to giant resonances, the small-amplitude approximation 
is often insufficient for low-frequency collective modes. 
This is especially the case for the quadrupole collective modes 
in open-shell nuclei. 
The oblate-prolate shape coexistences/mixings are typical examples.
It is also well known that the amplitude of the quadrupole vibration 
becomes very large in transient situations of the quantum phase transition 
from spherical to deformed,   
where the spherical mean field is barely stable or  
the spherical symmetry is broken only weakly. 
Many nuclei are situated in such a transitional region. 
It seems that this is one of the characteristic features of 
the quantum phase transition in the nucleus as a finite quantum many-body system.  
We need to go beyond the QRPA for describing 
such low-frequency quadrupole collective excitations.  
In view of the crucial role that the deformed shell structure and the pairing 
correlation play in generating the collectivity and determining the 
characters of these modes, 
it is desirable to construct a microscopic theory of LACM 
as an extension of the QRPA keeping its merit of deriving the collective 
coordinates from the huge number of microscopic degrees of freedom. 
Another important merit of the QRPA is that it is a quantum theory 
derived also by a new Tamm-Dancoff approximation in quantum field theory.  
Because we aim at constructing, on the basis of the TDHFB picture, 
a quantum theory of LACM capable of describing quantum spectra, 
it may be imperative, for justification of quantization of the collective coordinate,  
to formulate the quantum theory in such a way that it reduces to the QRPA 
in the small amplitude limit.  
In the next section, we briefly review various attempts toward this goal. 
 
\section{Problems in microscopic theories of collective motions}

\subsection{Boson expansion method}

One of the microscopic approaches to treat non-linear vibrations is 
the boson expansion methods \cite{kle91b}.  
For describing the quadrupole vibrations in transitional nuclei, 
the collective QRPA normal modes at the spherical shape are represented as 
boson operators and anharmonic effects ignored in the QRPA are evaluated 
in terms of a power series of the boson creation and annihilation operators. 
The boson expansion methods have been widely used for the investigation of
low-frequency quadrupole collective phenomena.   
This approach is perturbative in the sense that the microscopic structures of 
the collective coordinates and momenta are fixed at the spherical shape 
and non-linear effects are evaluated by a power series expansion 
in terms of these collective variables. 
To treat situations where the collective vibrations 
become increasingly larger amplitude 
and the non-linear effects grow to such a degree that
the microscopic structures of the collective variables themselves 
may change during the vibrational motion, 
it is desirable to develop a microscopic theory 
which can treat the non-linear effects in a non-perturbative way.  

\subsection{Generator coordinate method (GCM)}

In the application of the GCM to quadrupole collective phenomena, 
quantum eigenstates are described as superpositions of 
mean-field (generalized Slater determinant) states parametrized by the generator coordinates. 
This microscopic approach is widely used in conjunction with the angular momentum 
and number projections \cite{ben03}.  
A long-standing open problem in the GCM is the reliability of 
the collective masses (inertia functions)  evaluated 
with the real generator coordinates.  
For the center of mass motion, complex generator coordinates 
are needed to reproduce the correct mass, indicating that we have to explicitly treat 
the collective momenta in addition to the coordinates \cite{rin80}.
Another important problem in the GCM is the choice of the generator coordinates. 
Holzwarth and Yukawa \cite{hol74} once tried to find an optimal collective path 
by variationally determining the generator coordinate. This work  
stimulated the attempts to construct the microscopic theory of LACM. 

\subsection{Time-dependent HF (TDHF) method}

Needless to say, the TDHF is a powerful tool to microscopically describe the  
LACM taking place in heavy-ion collisions \cite{neg82}. 
The TDHF is insufficient for the description of quantum spectra of low-lying states,  
however, because of its semi-classical feature.  
On the other hand, as emphasized in section 3.6,  
its small-amplitude approximation, the RPA, can be formulated as 
a quantum theory and gives us a physical insights 
how the collective modes are generated as coherent superpositions of 
a large number of particle-hole excitations. 
It is thus desirable to extend the idea of deriving the RPA from the TDHF  
to large amplitude motions. 

\subsection{Adiabatic TDHF (ATDHF) method}

Challenges for constructing microscopic theory of LACM can be traced back to 
the pioneering works by Belyaev \cite{bel65}, Baranger and Kumar \cite{bar65} 
in which the collective potential and collective masses (inertial functions) 
appearing in the quadrupole collective Hamiltonian of Bohr and Mottelson are 
microscopically calculated on the basis of the time-dependent mean-field picture 
and with the use of the pairing-plus-quadrupole force \cite{bes69}. 
In the ATDHF theories, developed by Baranger and V\'en\'eroni \cite{bar78}, 
Brink \etal \cite{bri76}, Goeke and Reinhard \cite{goe78},   
more general schemes applicable to general effective interactions are given.  
There, under the assumption that the LACM is slow,  
the time-dependent density matrix is expanded in terms of the collective momentum,  
and the collective coordinate is introduced as a parameter 
describing the time dependence of the density matrix.  
Another ATDHF theory by Villars \cite{vil77} resembles the above approaches,  
but it is more ambitious in that it provides a set of equations to 
self-consistently determine the collective coordinates. 
It turned out, however, that these equations are insufficient to uniquely 
determine the collective coordinates. 
This problem was solved by properly treating the second order equation
(with respect to the collective momentum)
in the time-dependent variational principle 
(see  Mukherjee and Pal \cite {muk81} and the series of papers by   
Klein, Do Dang, Bulgac and Walet, reviewed in \cite{kle91a}).
It was also clarified through these works that 
the optimal collective path maximally decoupled from non-collective 
degrees of freedom coincides in a very good approximation 
with the valley in the large-dimensional configuration space 
associated with the TDHF states.  

\subsection{Self-consistent collective coordinate (SCC) method}

Attempts to construct the LACM theory  without assuming the adiabaticity 
and treating the collective coordinates and momenta on the same footing 
were initiated by Rowe \cite{row76} and Marumori \cite{mar77}. 
The problems remaining in these early works were solved in  
the SCC method formulated by Marumori \etal \cite{mar80}. 
The major aim of this approach is to 
extract the optimal collective path (collective submanifold), 
maximally decoupled from non-collective degrees of freedom,   
in the TDHF phase space of large dimension.
The collective submanifold is a geometrical object 
independent of the choice of the canonical coordinate system. 
This idea was developed also by Rowe \cite{row82},  
Yamamura and Kuriyama \cite{yam87}. 
These works yield a new insight into the fundamental concepts of collective motion. 
The SCC method was extended \cite{mat86} to include the pairing correlations 
and applied to anharmonic quadrupole vibrations \cite{mat85a,mat85b,yam93}. 
In these works, a perturbative method of solving its basic equations 
was adopted and 
it remained as an open task to develop a non-perturbative method 
of solution for genuine LACM. 
This task was attained by the adiabatic SCC (ASCC) method  
\cite{mat00}, in which the basic equations of the SCC method is solved 
using an expansion with respect to the collective momentum. 
Therefore the method is applicable to 
the change of the system in a wide range of the collective coordinate.  
This new method may also be regarded as 
a modern version of the ATDHF method initiated by Villars \cite{vil77}.  
Another approach similar to the ASCC method has been developed also 
by Almehed and Walet \cite{alm04} 
although the method of restoring the gauge invariance (number conservation) 
broken in the TDHFB states is not given there.  
In the next section, we briefly summarize the basic ideas of 
the microscopic theory of LACM along the lines of the ASCC method 
which is formulated respecting the gauge invariance.  

\section{Basic concepts of the microscopic theory of LACM}

\subsection{Extraction of collective submanifold}

As mentioned in the previous sections, 
the TDHFB theory can be formulated as a Hamilton dynamical system:  
the dimension of the phase space is quite large  
(twice of the number of all possible two quasiparticle states). 
The major aim of the microscopic theory of LACM is to extract 
the collective submanifold  from the TDHFB space, 
which is describable in terms of a few numbers of collective coordinates and momenta.
The collective Hamiltonian is then derived and requantized yielding the 
collective Schr{\"o}dinger equation. 

\subsection{Basic equations of the microscopic theory of LACM}

Let us assume that the time development of the TDHFB state  
is describable in terms of the single collective coordinate $q(t)$ and momentum $p(t)$. 
To describe the superfluidity, we need to introduce also the number 
variable $n(t)$ and the gauge angle $\varphi(t)$ conjugate to it 
（for both protons and neutrons in applications to nuclei).   
We assume that the TDHFB state can be written in the following form:  
\begin{eqnarray}
\ket{\phi(q,p,\varphi,n)} &= e^{-i \varphi\Nt}\ket{\phi(q,p,n)},\\
\ket{\phi(q,p,n)} &= e^{ip\Qhat(q) + in\That(q)}\ket{\phi(q)}.
\label{eq:TDHFBstate}
\end{eqnarray}
Here $\ket{\phi(q,p,n)}$ is an intrinsic state for  
the pairing rotational degree of freedom parametrized by $\varphi$, 
$\ket{\phi(q)}$ represents a non-equilibrium HFB state called 
{\it moving-frame HFB state}, 
$\Qhat(q)$ and $\That(q)$ are one-body operators 
called {\it infinitesimal generators}, 
and $\Nt$ and $n$ are defined by  
$\Nt \equiv \Nhat- \bra{\phi(q)}\Nhat\ket{\phi(q)} \equiv \Nhat-N_0$ and
$n  \equiv \bra{\phi(q,p,n)}\Nhat\ket{\phi(q,p,n)}- N_0 \equiv  N - N_0$, 
$\Nhat$ being the number operator. 

We determine the microscopic structure of the 
infinitesimal generators $\Qhat(q), \That(q)$ and 
the moving-frame HFB state $\ket{\phi(q)}$
on the basis of the time-dependent variational principle:  
\begin{eqnarray}
\delta \bra{\phi(q,p,\varphi,n)} i\frac{\del}{\del t} - \Hhat \ket{\phi(q,p,\varphi,n)}  = 0,
\label{eq:TDVP}
\end{eqnarray}
where $\Hhat$ is a microscopic many-body Hamiltonian.
Expanding in powers of $p$ and $n$ and keeping terms up to the second order 
in $p$, we obtain  
{\it the moving-frame HFB equation}
\begin{eqnarray}
 \delta\bra{\phi(q)}\Hhat_M(q)\ket{\phi(q)} = 0, 
\label{eq:mfHFB}
\end{eqnarray}
where $\Hhat_M(q)$ is the moving-frame Hamiltonian defined by  
\begin{eqnarray}
\Hhat_M(q) = \Hhat - \lambda(q) \Nt - \frac{\del V}{\del q}\Qhat(q),  
\end{eqnarray}
and {\it the moving-frame QRPA equations}（also called {\it local harmonic equations}）
\begin{eqnarray}
 \delta\bra{\phi(q)}[\Hhat_M(q), \Qhat(q)] - \frac{1}{i} B(q)\Phat(q)
 \ket{\phi(q)}  = 0, 
\label{eq:mfQRPA1}
\end{eqnarray}
\begin{eqnarray}
 \delta\bra{\phi(q)} & [\Hhat_M(q),\frac{1}{i}\Phat(q)] -
 C(q)\Qhat(q) \nonumber \\
  &- \frac{1}{2B(q)} \left[ \left[
  \Hhat_M(q), \frac{\del V}{\del q}\Qhat(q)
\right], \Qhat(q) \right] \nonumber \\
  &- \frac{\del
 \lambda}{\del q} \Nt \ket{\phi(q)} = 0,
\label{eq:mfQRPA2}
\end{eqnarray}
where $\Phat(q)$ is the displacement operator defined by 
\begin{eqnarray}
 \ket{\phi(q + \delta q)} = 
e^{-i \delta q \Phat(q)}\ket{\phi(q)} ,
\end{eqnarray}
and 
\begin{eqnarray}
 C(q) = \frac{\del^2 V}{\del q^2}
 + \frac{1}{2B(q)}\frac{\del B}{\del q}\frac{\del V}{\del q}. \label{eq:defC}
\end{eqnarray}
The quantities $C(q)$ and $B(q)$ are related to 
the eigenfrequency $\omega(q)$ of the local normal mode 
obtained by solving the moving-frame QRPA equations 
through $\omega^2(q)=B(q)C(q)$. 
Note that these equations are valid also for regions with negative curvature 
($C(q)<0$) where $\omega(q)$ takes an imaginary value. 
The double commutator term in (\ref{eq:mfQRPA2})
stems from the $q$ derivative of $\Qhat(q)$ and 
represents the curvature of the collective path. 

Solving the above set of equations, the microscopic structure of 
the infinitesimal generators, $\Qhat(q)$ and $\Phat(q)$, are determined:  
they are explicitly expressed as bilinear superpositions of 
the quasiparticle creation and annihilation operators locally defined 
with respect to the moving-frame HFB state $\ket{\phi(q)}$. 
The collective Hamiltonian is given by 
\begin{eqnarray}
 \Hc(q,p,n) =\bra{\phi(q,p,n)}\Hhat\ket{\phi(q,p,n)} 
 =V(q) + \frac{1}{2}B(q)p^2 + \lambda(q)n,
\label{eq:collH}
\end{eqnarray}
where $V(q)$, $B(q)$ and $\lambda(q)$ represent the collective potential, 
inverse of the collective mass and the chemical potential, respectively.  
Note that they are functions of the collective coordinate $q$. 
The basic equations, (\ref{eq:mfHFB}), (\ref{eq:mfQRPA1}) and (\ref{eq:mfQRPA2}), 
reduce to the well-known HFB and QRPA equations 
at the equilibrium points where $\del V/\del q=0$. 
Thus,  the LACM theory outlined above is a natural extension of 
the HFB-QRPA theory to non-equilibrium states. 

\subsection{Relation to the constrained HFB approach} 

The moving-frame HFB equation (\ref{eq:mfHFB}) looks like
the constrained HFB (CHFB) equation, 
but it is essentially different from the CHFB in that 
the infinitesimal generator $\Qhat(q)$ (corresponding to the constraint operator) 
is self-consistently determined together with $\Phat(q)$ 
as a solution of the moving-frame QRPA equations,  
(\ref{eq:mfQRPA1}) and (\ref{eq:mfQRPA2}), 
locally at every point of the collective coordinate $q$. 
Therefore, unlike the constraint operator in the CHFB method, 
its microscopic structure changes as a function of $q$.
In other words, the LACM takes place  
choosing the locally optimal `constraint operator' at every point of $q$
along the collective path.  
When the LACM of interest is described by more than one collective coordinates, 
we try to extract the collective hypersurface embedded in the large-dimensional 
TDHFB space by extending the above equations to the multi-dimensional cases and 
derive the collective Schr{\"o}dinger equation by requantization. 
This attempt has some features in common with the problem of 
{\it quantization of constrained dynamical system} \cite{row82}. 
It is interesting to discuss the microscopic foundation 
of the so-called Pauli prescription 
(frequently used in quantizing phenomenological collective Hamiltonians)
from this point of view.  In discussing this issue, it is important to note 
that the LACM is constrained on the collective hypersurface 
not by external constraining forces but by the dynamics of itself. 
Namely, the collective hypersurface is generated as a consequence 
of the dynamics of the quantum many-body system under consideration. 
It seems that this is a unique and quite attractive feature of the subject 
under discussion. 

\subsection{Gauge invariance with respect to the pairing rotational angle} 

An important problem in formulating the microscopic LACM theory 
based on the TDHFB approximation is how to respect the number conservation. 
As is well known, one of the merits of the QRPA is that the zero-frequency 
Nambu-Goldstone mode restoring the number conservation  
is decoupled from other normal modes of vibration \cite{bri05}. 
Our problem is how to generalize this concept to non-equilibrium HFB states.   
A clue for solving this problem is obtained by noting that  
the basic equations, 
(\ref{eq:mfHFB}), (\ref{eq:mfQRPA1}) and (\ref{eq:mfQRPA2}),
are invariant against rotations of the gauge angle $\varphi$
at every point of $q$. 
Actually, we can determine the infinitesimal generator $\That(q)$
associated with the pairing-rotational degree of freedom
in the same way as $\Qhat(q)$ and $\Phat(q)$,  
and obtain the pairing rotational energy proportional to $n^2$ 
as an additional term to the collective Hamiltonian (\ref{eq:collH}). 
Although this term vanishes by setting $N = N_0$
to respect the number conservation, 
the consideration of the gauge invariance is essential for 
correctly treating the LACM in systems with superfluidity. 
Specifically, we need to set up a gauge fixing condition in 
practical calculations (see \cite{hin07} for details). 
 
\section{Open problems in the microscopic theory of LACM}
 
\subsection{Contributions of the time-odd mean field to the collective mass}

In the microscopic calculation of the collective mass (inertial function), 
the cranking mass is widely used. 
It is obtained through the adiabatic perturbation treatment of
the time development of the mean field.  
As stressed by Belyaev \cite{bel65}, Baranger and V\'en\'eroni \cite{bar78}, 
the effects of the time-odd components (breaking the time-reversal invariance) 
induced by the time evolution of the mean field are ignored in the cranking mass. 
These effects are self-consistently taken into account in the 
ASCC collective mass obtained by solving 
Equations~(\ref{eq:mfQRPA1}) and (\ref{eq:mfQRPA2}). 
On the other hand, one can derive the collective Schr{\"o}dinger equation 
from the GCM equation by making the Gaussian overlap approximation (GOA).   
It is not clear, however,  to which extent the time-odd mean-field effects are 
taken into account in the GCM-GOA mass evaluated in this way with 
real generator coordinates.  
In view of the importance of the collective mass in the 
dynamics of LACM, the time-odd mean-field effects have been studied 
extensively (see e.g. \cite{dob95}).  
In spite of these efforts, it may be fair to say that 
we are still far from a full understanding of the time-odd effects.  
Thus, it remains as a great subject for future to evaluate the time-odd mean-field 
effects on the collective mass using energy density functionals 
currently under active development.    

\subsection{Different meaning of `adiabatic'}

There is another important difference 
between the ASCC collective mass and the cranking mass.    
Let us first note that the meaning of the adjective `adiabatic' 
in the ASCC method is different from that of the adiabatic perturbation theory.   
In the ASCC method, it just means that the collective kinetic energy term 
higher than the second order in the power-series expansion with respect to the 
collective momentum is omitted and, in contrast to the adiabatic perturbation, 
the smallness of the collective kinetic energy in comparison with
the intrinsic two-quasiparticle excitation energies is not indispensable. 
Recalling that the ASCC collective mass coincides with the QRPA collective mass 
at the HFB equilibrium point, 
one can easily confirm this difference by considering 
the spherical QRPA limit in the pairing-plus-quadrupole force model,  
where  the time-odd mean-field effect is absent. 
At the spherical HFB equilibrium point, in fact, the QRPA collective mass 
reduces to the cranking mass in the limit that the frequency of the QRPA  normal mode 
vanishes \cite{bes69}. 
In this connection, it may be pertinent to emphasize the difference between  
the ASCC collective mass and the ATDHFB collective mass of Baranger and V\'en\'eroni.  
Specifically, the former is determined by the local QRPA mode 
and reduces to the QRPA collective mass at the HFB equilibrium point, 
while the latter is related to the cubic inverse energy-weighted sum rule \cite{gia80}
and does not reduce to the QRPA collective mass in this limit.  
Therefore, it is also interesting to make a systematic comparison between 
different collective masses including the ASCC, cranking, GCM-GOA 
and ATDHFB collective masses. 

\subsection{Deeper understanding of the pair-hopping mechanism}

The collective mass represents the inertia of the many-body system 
against an infinitesimal change of the collective coordinate $q$
during the time evolution of the mean field. 
It is a local quantity and varies as a function of $q$. 
What is the microscopic mechanism of determining the collective mass ?  
This is one of the central questions in our study of many-body dynamics 
of the LACM. Concerning this question, it is well known that 
the pairing correlation plays a crucial role. 
Because the single-particle-energy spectrum in the mean field changes 
as a function of $q$, the level crossing at the Fermi energy successively occurs 
during the LACM.  In the presence of the pairing correlation,  
the many-body system can easily rearrange to take the lowest-energy  
configurations at every value of $q$, i.e. the system can easily change $q$.    
The easiness/hardness of the configuration rearrangements at the 
level crossings determines the adiabaticity/diabaticity of the system. 
Since the inertia represents a property of the system trying 
to keep a definite configuration, we expect that  
the stronger the pairing correlation, the smaller the collective mass.  
The nucleon-pair hopping mechanism at the successive level crossings 
at the Fermi surface is modeled by Barranco \etal \cite{bar90}. 
It yields the collective mass,  called hopping mass,  
which has been applied to the exotic decays and the tunneling phenomena 
between the superdeformed (SD) and normal deformed states \cite{bar90,yos01}. 
Of course, smooth changes of single-particle wave functions as functions of $q$ 
also contribute to the collective mass in addition to the configuration changes. 
To deepen our understanding of the collective mass of LACM, 
it is desirable to make a comprehensive analysis of the microscopic mechanism 
determining it.  A systematic comparison between the hopping mass 
and other collective masses discussed above will certainly serve for this purpose.   

\subsection{Application to the shape coexistence/mixing phenomena}

As discussed in section 2, one of the most interesting LACM phenomena is the 
oblate-prolate shape coexistence/mixing in the proton-rich $^{68}$Se and $^{72}$Kr 
region.  Quite recently, we have applied the ASCC method to them and successfully  
determined the collective path which runs through the triaxially deformed valley 
and connects the oblate and the prolate HFB minima. 
Evaluating the rotational moments of inertia on the collective path 
and requantizing the collective Hamiltonian, 
we have derived the collective Schr{\"o}dinger equation describing the coupled motion 
of the large-amplitude shape vibrations and the three-dimensional (3D) rotational motions  
\cite{hin08,hin09}. We have thus found a number of interesting features which change 
when going from nucleus to nucleus. 
For instance, the $^{68}$Se nucleus exhibits intermediate features between the 
oblate-prolate shape coexistence and the $\gamma$-soft rotors, while 
in the neighboring $N=Z$ nucleus $^{72}$Kr, 
we have found that the localization of the collective wave function 
in the $(\beta,\gamma)$ plane significantly develops 
with increasing rotational angular momentum.     
It is certainly desirable to carry out this kind of microscopic analysis for 
a wide variety of shape coexistence/mixing  phenomena. 
We believe that theoretical and experimental investigations of these phenomena 
will be very fruitful and bring about plenty of new ideas 
on nuclear structure and dynamics.

\begin{figure}
\includegraphics[width=1.0\textwidth]{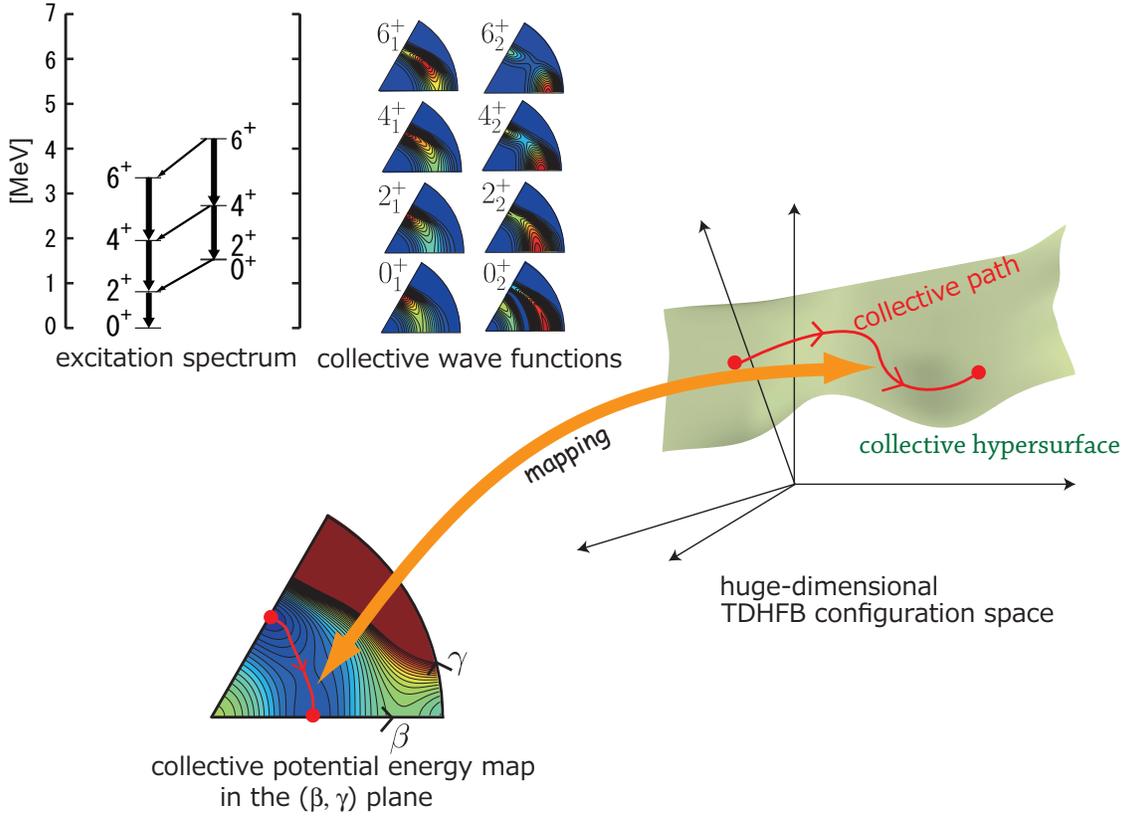}
\caption{  
Illustration of basic concepts of LACM. 
The collective path and the collective hypersurface embedded in 
the huge-dimensional TDHFB configuration space ({\it right-hand side}).   
Mapping of the collective path and the hypersurface
into the $(\beta,\gamma)$ plane and the collective potential energy 
on it ({\it lower part in the left-hand side}). 
The excitation spectrum and collective wave functions obtained 
by solving the collective Schr{\"o}dinger equation ({\it upper part in the left-hand side}).
In this illustration, the result of a microscopic calculation for the oblate-prolate shape 
coexistence/mixing phenomenon in $^{68}$Se is used,  
where the collective path is self-consistently determined by 
solving the ASCC equations while the collective potential 
and the collective masses are evaluated by solving the CHFB 
and the moving-frame QRPA equations, respectively, 
with the pairing-plus-quadrupole force 
(the quadrupole pairing is also taken into account).  
This calculation may be regarded as a first step 
toward a fully self-consistent microscopic derivations of the 5D quadrupole 
collective Hamiltonian starting from modern density functionals. 
}
\label{fig:illust}
\end{figure}

\subsection{Microscopic derivation of the Bohr-Mottelson collective Hamiltonian}

Extending the one-dimensional (1D) collective path to the two-dimensional 
(2D) hypersurface and map it on the $(\beta,\gamma)$ plane, 
we shall be able to microscopically derive the 
five-dimensional (5D) quadrupole collective Hamiltonian of 
Bohr and Mottelson \cite{boh75}.
In this derivation, the moments of inertia of the 3D rotation 
may be evaluated at every point of the 2D hypersurface generalizing 
the Thouless-Valatin equations to those at non-equilibrium points. 
The microscopic derivation of the Bohr-Mottelson collective Hamiltonian
is a well-known long-standing subject in nuclear structure physics, 
but we are still on the way to the goal
(see \cite{ben08,nik09} for examples of recent works and \cite{pro09} for a review). 
We illustrate in \fref{fig:illust} the basic concepts of the microscopic LACM theory 
and  the result of a recent calculation.  
 
\subsection{Extension to high-spin states}

In the above approach, variations of intrinsic structure due to rotation are not taken into 
account. Therefore, the range of its applicability is limited to low-spin states. 
A promising way of constructing LACM theory applicable to high-spin states 
is to adopt the rotating mean-field picture. 
Specifically, it is interesting to develop LACM theory 
on the basis of the HFB approximation in a rotating frame    
(it is possible to formulate the SCC method in a form suitable for treating the 
rotational motions \cite{shi01,kan94}). 
Such an approach was once tried in \cite{alm04}.    
It remains as a future challenge to construct microscopic theory 
of LACM at high spin, which is capable of self-consistently taking into account
variation of intrinsic structure due to rapid rotation. 

\subsection{Combining with better density functionals}

As seen in a number of contributions to this special issue on open problems 
in nuclear structure, very active works are going on to build  
a universal nuclear energy density functional.  
It is certainly a great challenge to make a systematic microscopic calculation 
for LACM phenomena using better energy density functionals. 
For carrying out such ambitious calculations, it is certainly necessary 
to develop efficient numerical algorithms to solve the basic equations 
of the LACM theory.  In practical applications, for instance, 
we need to iteratively solve the moving-frame HFB equation 
and the moving-frame QRPA equations at every point on the collective path.
When we extend these equations to 2D hypersurfaces, 
the numerical calculation grows to a large scale.  
Especially, an efficient method of solving the moving-frame QRPA 
equations is needed.  
An extension of the finite amplitude method \cite{nak07} 
into a form suitable for this purpose may be promising. 
It may also be worthwhile to examine the applicability 
of the separable approximation \cite{kle08} to the effective interaction 
derived from the energy density functionals.     

\subsection{LACM in odd-$A$ nuclei}

So far we have limited our discussions to LACM in doubly even nuclei only.  
In fact, microscopic description of LACM in odd-$A$ nuclei remains 
as a vast unexplored field. 
Needless to say, unified treatment of the seemingly contradictory 
concepts of single-particle and collective modes of motion is 
the central theme of the phenomenological Bohr-Mottelson model 
of nuclear structure.  
Low-lying states in odd-$A$ nuclei provide a wealth of data  
exhibiting their interplay. 
In view of the great success of the Bohr-Mottelson approach, 
it is extremely important to develop a microscopic theory 
capable of treating the single-particle and collective modes 
in a unified manner. 
In fact, various microscopic theories of particle-vibration coupling 
have been developed: e.g. the nuclear field theory \cite{bor77} 
and the boson-expansion method for odd-$A$ nuclei \cite{kle91b}. 
These available theories treat the particle-vibration couplings 
in a perturbative manner starting from the small-amplitude approximation 
(RPA/QRPA) for collective excitations at an equilibrium point of the mean field.  
Non-perturbative method capable of treating LACM in odd-$A$ nuclei 
(generally speaking, LACM in the presence of several quasiparticles) 
is lacking, however.    
This is an extremely difficult but challenging subject for future. 

\section{Large-amplitude collective phenomena at high spin}

The nucleus exhibits a rich variety of non-linear collective phenomena 
awaiting applications of the microscopic LACM theory. 
Certainly, microscopic description of spontaneous fission 
from the viewpoint of non-linear/non-equilibrium physics 
is one of the major goals. 
Another vast unexplored field is the microscopic study of 
LACM at finite temperature. 
Microscopic mechanism of damping and dissipation of 
various kinds of LACM is a long-term subject.  
It seems particularly interesting to explore both theoretically 
and experimentally how the character of LACM changes
when going from the yrast  to the compound-nucleus regions.   
These subjects are outside the scope of the present article, however. 
Restricting our scope to the LACM at zero temperature, 
in this section, we discuss a few open problems in 
rapidly rotating nuclei (see \cite{sat05} for a review on high-spin states).

\subsection{Tunneling decay of superdeformed (SD) states and high-K isomers}

It is quite interesting to apply the microscopic LACM theory to 
the quantum tunneling phenomena from the SD states 
(with axis ratio about 2:1) to the compound-nucleus states \cite{yos01}, 
because the tunneling probability depends quite sensitively on the collective mass 
and the collective path connecting the initial and final states. 
The tunneling decay from high-$K$ to low-$K$ isomers is also interesting 
\cite{nar96}, because it poses a unique question as to  
which of the two competing decay paths dominates; 
one through LACM in the triaxial shape degree of freedom and 
the other in the orientation degree of freedom of the angular momentum vector 
(with respect to the principal axes of the body-fixed frame).  

\subsection{Large-amplitude wobbling motions and chiral vibrations}

A new mode of 3D rotation associated with the spontaneous breaking of 
the axial symmetry is called wobbling motion; it is describable 
as a boson (small-amplitude vibration-like) excitation from the yrast line 
\cite{boh75,sho09}.  The observed rotational band associated with 
the double excitations of the wobbling mode
indicates, however,  the presence of significant anharmonicity \cite{jen02}. 
It is an interesting open problem to explore what will happen when the 
amplitude of the wobbling mode increases and the uniformly rotating 
nucleus becomes unstable against this collective rotational degree of freedom 
\cite{mat04b,olb04}. Another interesting issue concerning new modes of 
3D rotation is the possibility of doublet rotational bands called 
chiral band \cite{fra01,koi04,muk07}. 
In triaxially deformed nucleus, one can define {\it chirality} in terms of 
the directions of the collective rotational angular momentum 
and those of the quasiparticle angular momenta of both protons and neutrons. 
When the rotating HFB mean field involves such an intrinsic structure, 
the two solutions corresponding to the right-handed and left-handed configurations 
are degenerate.  Then, a chiral doublet pattern is expected to appear in the 
rotational band structure.  In a transient situation, where the barrier 
separating the two HFB solutions is still in an early stage of development, 
the large-amplitude vibrations connecting the two configurations, called 
chiral vibrations, may occur \cite{fra01,koi04,muk07}.
It remains for the future to apply the microscopic LACM theory 
to these phenomena unique to rapidly rotating nuclei with triaxial shape.  

\subsection{Large-amplitude vibrations associated with the reflection symmetry breaking}

Rich experimental data exhibiting the parity-doublet pattern are available,  
indicating that the reflection symmetry is broken in their mean fields \cite{but96}. 
The energy splitting of the doublet changes as a function of 
rotational angular momentum \cite{jol99}. 
Thus, it is quite interesting to investigate,  
on the basis of the microscopic LACM theory, 
how the quantum tunneling motion 
between the left- and right-configurations is affected by rapid rotation.     
Also interesting in this connection is the recent observation of alternating 
parity bands \cite{rev06,fra08} 
that exhibit a transitional feature toward the static octupole deformation. 
A considerable number of SD states are expected to be very soft 
with respect to the shape vibrational degrees of freedom 
simultaneously breaking the reflection and axial symmetries.
In fact, the soft octupole vibrations built on the SD yrast states 
have been observed \cite{nak96,ros01}.
When the SD mean field becomes unstable against this kind of vibrations,
a new class of SD states having exotic shapes (like banana) may appear 
\cite{nak92,ina02}. 
In transitional situations, the large-amplitude vibrations 
associated with the instability toward such exotic shapes may take place.   
Possible appearance of exotic shapes is not restricted to SD high-spin states. 
For example, the symmetry-unrestricted HFB calculation \cite{yam01} 
yields a local minimum with tetrahedral shape near the ground state of $^{80}$Zr.   
The potential energy surface is shallow, however, indicating that 
we need to take into account the large-amplitude tetrahedral shape fluctuation.
Generally speaking, microscopic LACM theories are required for describing  
collective motions in transitional regions of quantum phase transition,  
where some symmetry is weakly broken or tends to be broken.

\section{Concluding remarks}

One of the fundamental questions of nuclear structure physics is 
why and how a variety of LACM emerges in consequence of 
quantum many-body dynamics. The nucleus provides us valuable 
opportunities to make a detailed study of the microscopic dynamics 
generating the collectivities.     
The microscopic derivation of the quadrupole collective Hamiltonian 
started more than half a century ago, but the challenge to 
construct a fully self-consistent microscopic theory of LACM 
has encountered a number of serious difficulties.  
Finally, however, these long-term efforts have yielded the new concept 
of {\it collective submanifold} and a deeper understanding of 
what is {\it collectivity}. At the same time, new efficient methods  
of numerical calculation are now under active development. 
Thus, in the coming years, fruitful applications 
to low-frequency collective phenomena are envisaged. 
This means that a new era in the microscopic study of 
nuclear collective dynamics is opening.

\ack
This work is supported by the Grant-in-Aid for Scientific Research(B)
(No. 21340073). One of the authors (N.H.) is supported by the Special Postdoctoral
Researcher Program of RIKEN.

\end{document}